\documentclass[prb,twocolumn,showpacs]{revtex4}
\usepackage{epsfig}
\usepackage{graphics}
\usepackage{amsfonts}
\usepackage{amsmath,amssymb}

\begin{document}

\title{Confined One Dimensional Harmonic Oscillator as a Two-Mode System}
\author{V. G. Gueorguiev$^{1,2}$\footnote{On leave of absence from Institute of
Nuclear Research and Nuclear Energy, Bulgarian Academy of Sciences,
Sofia 1784, Bulgaria.}, A. R. P. Rau$^{2}$, and J. P. Draayer$^{2}$}
\address{$^{1}$Lawrence Livermore National Laboratory, Livermore, CA 94550\\
$^{2}$Department of Physics and Astronomy, Louisiana State
University, Baton Rouge, LA 70803 }

\pacs{03.65.-w, 02.70.-c, 02.60.-x, 03.65.Ge}

\begin{abstract}
The one-dimensional harmonic oscillator in a box problem is possibly the
simplest example of a two-mode system. This system has two exactly
solvable limits, the harmonic oscillator and a particle in
a (one-dimensional) box. Each of the two limits has a characteristic spectral structure 
describing the two different excitation modes of the system. 
Near each of these limits, one can use perturbation theory to achieve an
accurate description of the eigenstates. Away from the exact
limits, however, one has to carry out a matrix diagonalization because
the basis-state mixing that occurs is typically too large to be reproduced
in any other way. An alternative to casting the problem in terms of one or
the other basis set consists of using an ``oblique" basis that uses
both sets. Through a study of this alternative in this one-dimensional problem,
we are able to illustrate practical solutions and infer the applicability
of the concept for more complex systems, such as in the study of
complex nuclei where oblique-basis calculations have been successful.
\newline{\newline
Keywords: one-dimensional harmonic oscillator, particle in a box, exactly
solvable models, two-mode system, oblique basis states, perturbation
theory, coherent states, adiabatic mixing}
\end{abstract}
\maketitle

\section{Introduction}

\quad The understanding of a physical system is closely linked to how
well one can determine its eigenstates. Typically a set of basis states
that works well in one limit, fails in another. And more general methods,
such as variational schemes, perturbation theory, or fixed-basis matrix
diagonalizations typically begin with a reasonable Hamiltonian and some appropriate
set of basis states that yield
a good description of the system.

When applying perturbation theory, one is usually concerned with a
small perturbation of an exactly solvable Hamiltonian system.
However, there are many examples when the Hamiltonian has more than a
single exactly solvable limit. This is a common situation when a
dynamical symmetry group is used in the construction of the Hamiltonian.
A simple example is the hydrogen atom in an external magnetic field:
With  increasing field strength, particularly for magnetic fields
exceeding  a so-called critical value of $2.35 \times 10^5$ T, the
system  changes from the spherical symmetry of the Coulomb problem to
the  cylindrical symmetry of the diamagnetic Hamiltonian.\cite{Rau-New
patterns} Another example that occurs in a variety of condensed matter
problems is a particle confined to two dimensions in an external  magnetic
field.\cite{Rosas et al-2000} In nuclear physics, the  Interacting Boson
Model classifies many nuclei according to one of three dynamical
symmetries.\cite{IBM}  {\it But, what should be done if the system is
nowhere near any of the  exact limits?} In these situations, the problem
may be approached better by using states associated with all the appropriate
nearby limits. This set of states will form an ``oblique" (mixed-mode)
basis for the calculation.\cite{VGG 24MgObliqueCalculations, VGG PhD} In
general, such a basis is non-orthogonal and may even be over complete.
Nevertheless, as recent studies demonstrate, such oblique bases have
merit. In this paper, we use a  pedagogically simple problem to
illustrate the oblique basis approach.

The one-dimensional harmonic oscillator in a one-dimensional box has been
used to illustrate different aspects of mixing generated by
two interactions. Barton, Bray, and Mackane used the model to study the
effects of distant boundaries on the energy levels of a
one-dimensional quantum system.\cite{Barton-Bray-Mckane-1990}  Studies
have also been done for the cylindrically symmetric system of a
three-dimensional harmonic oscillator between two impenetrable
walls.\cite {Marin and Cruz-1988} However, these studies did not discuss
the bi-modal nature of the problem. Some authors have generalized the
one-dimensional harmonic oscillator problem by introducing time-dependent
parameters in the Hamiltonian\cite{Lejarreta-1999} to pass between
the two limiting cases of a free particle and the harmonic oscillator
solutions. The infinite square well and the harmonic oscillator have been
considered as two limiting cases of a power-law potential within the
context of wave packet collapses and revivals.\cite {Robinett-2000
AJP,Robinett-2000 JMP} Recent research in modified uncertainty relations
has also shown related and interesting  behavior.\cite{modified
uncertainty relations}

In this paper, we demonstrate the concept of an oblique basis approach
by considering the simple two-mode system of the one-dimensional harmonic
oscillator in a box.  First, in Sec.~\ref{HOin1Dbox}, we discuss the
concept as well as the exactly solvable limits of this toy model.
A qualitative discussion of the expected spectrum of the one-dimensional
harmonic oscillator in a box is given in Sec.~\ref{Spectral Structure},
along with an example spectrum and quantitative estimates.  In
Sec.~\ref{Hilbert space} some specific problems related to the structure
of the Hilbert space are addressed.  Sec.~\ref{Calculations and Results}
contains specific toy model calculations and results as well as a
discussion of a new interesting behavior, similar to that observed in
nuclear structure studies.\cite{VGG 24MgObliqueCalculations, VGG PhD} The
discussion in Sec.~\ref{Calculations and Results} is focused on a
quasi-perturbative behavior and a coherent structure within the ``strong
mixing" region\cite{Rau-strong mixing} where the system is far from any
exact limit. Our conclusions are given in Sec.~\ref{Conclusions} and suggested student exercises in Sec.~\ref{Student exercises}.

\section{Harmonic Oscillator in a One-Dimensional Box}
\label{HOin1Dbox}

\quad
Let us start with an abstract two-mode system. For simplicity,
we assume that the Hamiltonian of the system has two exactly solvable
limits:
\begin{equation}
H=(1-\lambda )H_{0}+\lambda H_{1}.
\label{2-mode system H}
\end{equation}
Clearly this is set up so that $H$ $\rightarrow H_{0}$ in the limit
$\lambda \rightarrow 0$ and $H$ $\rightarrow H_{1}$ when
$\lambda \rightarrow 1$. In the vicinity of these two limits, one can use
standard perturbation theory for one Hamiltonian perturbed by the other.\cite{Fernandez-2000}  Usually, somewhere in 
between these two limits there is a critical value of $\lambda$ that is related to 
the strongly mixed regime of the system. This value of $\lambda$ could be anywhere in the interval  $(0,1)$, a convenient choice being $\lambda \approx \frac{1}{2}$. Sometimes, a further symmetry breaking Hamiltonian $H_2$ can be explicitly introduced
by adding $\lambda (1-\lambda)H_2$ to $H$.

In Eq.~(\ref{2-mode system H}), the variable $\lambda $ has been
introduced to simplify the discussion. In general, there will be more
than just one such parameter in the Hamiltonian.\cite{Skyrme-1957 CinQM}
Often the exactly solvable limits are
described as hypersurfaces in the full parameter space. It could even be
that there are three or more exactly solvable limits. For example, the
Interacting Boson Model (IBM)\cite{IBM} for nuclear spectra has three
exactly solvable limits.\cite{MoshinskyBookOnHO,VarnaPaper} Another
example with three exactly solvable limits is the commonly used schematic
interaction with non-degenerate single-particle energies ($\varepsilon
_{i}$), pairing ($P^{+}P$)  two-body interactions,\cite{Dukelsky et
al-Pairing} and quadrupole-quadrupole\cite{Chairul Bahri's Thesis}
($Q\cdot Q$) two-body interactions:
\[
H=\varepsilon _{i}N_{i}-GP^{+}P-\chi Q\cdot Q.
\]

Here, we consider what is perhaps the simplest two-mode system that shares the essential features of such problems
while remaining pedagogically instructive. The Hamiltonian of a
one-dimensional harmonic oscillator in a one-dimensional box\cite{Armen
and Rau} of size
$2L$ has the form:
\begin{equation}
H=\frac{1}{2m}p^{2}+V_{L}(q)+\frac{m\omega ^{2}}{2}q^{2},
\label{H-ho-1Dbox}
\end{equation}
where $V_{L}(q)$ is the confining potential taking the value zero for $\left|
q\right| <L$ and $\infty $ for $\left| q\right| \geq L$, and $\omega$
is the oscillator frequency. This system has two exactly
solvable limits.

The first limit of the toy model in Eq.~(\ref{H-ho-1Dbox}) is $\omega
=0$ when it reduces to
a free particle in a one-dimensional box of
size $2L$,
\begin{equation}
H_{0}=\frac{1}{2m}p^{2}+V_{L}(q).
\label{1D box Hamiltonian}
\end{equation}
The eigenvectors and energies are labeled by $n=0,1,...$ and given by

\begin{eqnarray}
\Phi _{n}(q) &=&\left\{
\begin{tabular}{lll}
$\sqrt{\frac{1}{L}}\cos \left( (n+1)\frac{\pi }{2}\frac{q}{L}\right) $ & if
& $n$ is even \\
$\sqrt{\frac{1}{L}}\sin \left( (n+1)\frac{\pi }{2}\frac{q}{L}\right) $ & if
& $n$ is odd
\end{tabular}
\right. , \label{1D box FW and En} \\
E_{n} &=&\frac{1}{2m}\left( (n+1)\frac{\pi }{2}\right) ^{2}\left( \frac{
\hbar }{L}\right) ^{2}. \nonumber
\end{eqnarray}
This limit corresponds to extreme nuclear matter when the short range
nuclear force can be described as an effective interaction represented by
a square-well potential.\cite{Heyde-1994} We can think of this limit as
the one-dimensional equivalent of a three-dimensional model where
nucleons are confined within a finite volume of space representing the
nucleus. Recently such effective potentials for the Bohr Hamiltonian have
been used to introduce symmetries in the critical point of quantum phase
transitions.\cite{X(5)-E(5)}

The other exactly solvable limit of the toy model in
Eq.~(\ref{H-ho-1Dbox}), when $L \rightarrow \infty$, is the
harmonic oscillator in one dimension,
\begin{equation}
H_{1}=\frac{1}{2m}p^{2}+\frac{m\omega ^{2}}{2}q^{2}.
\label{Harmonic oscillator Hamiltonian}
\end{equation}
In dimensionless coordinates,
\[
q\rightarrow \tilde{q}\sqrt{\frac{\hbar }{m\omega }},\quad p\rightarrow
\tilde{p}\sqrt{m\hbar \omega },
\]
we have
\[
H_{1}=\hbar \omega \frac{1}{2}\left( \tilde{p}^{2}+\tilde{q}^{2}\right) .
\]
The eigenvectors and energies are labeled by $n=0,1,...$ and are given
by
\begin{eqnarray}
\Psi _{n}(q) &=&\sqrt{\frac{1}{b\, n!\, 2^{n}\sqrt{\pi }}}H_{n}\left(
\frac{q}{b}
\right) \exp \left( -\frac{1}{2}\frac{q^{2}}{b^{2}}\right),
\label{Harmonic oscillator WF and En} \\
E_{n} &=&\hbar \omega \left( n+\frac{1}{2}\right) ,
\quad b=\sqrt{\frac{\hbar }{m\omega }}, \nonumber
\end{eqnarray}
where $H_{n}$ are the Hermite polynomials. This limit is essentially
the harmonic oscillator model for nuclei.

In a one-dimensional toy model, the anharmonic oscillator with a quartic
anharmonicity would be the appropriate counterpart of the $Sp(6,R)$ shell
model\cite{Sp(6)models} since the quadrupole-quadrupole interaction $Q\cdot
Q$ goes as $\sim r^{4}$ and $Q$ connects harmonic oscillator
shells with like parity. If we restrict the model space to only one
harmonic oscillator
shell, then we can use the algebraic quadrupole moment $\tilde{Q}$ of
Elliott\cite{Elliott's SU(3) model} because within a single shell $\tilde{Q}
$ is the same as $Q$.\cite{MoshinskyBookOnHO} Thus, for single shell
studies, it is appropriate to consider the one-dimensional harmonic
oscillator as representative of the $SU\left( 3\right) $ shell model
for nuclei.

\section{Spectral Structure at Different Energy Scales}
\label{Spectral Structure}

\quad Often in physics the spectrum of a system has different
characteristics over different energy regimes. This usually reflects the
existence of different excitation modes of the system. For the toy model
Hamiltonian in  Eq.~(\ref{H-ho-1Dbox}), we can define three spectral
regions:

\begin{itemize}
\item Spectrum of a particle in a one-dimensional box as in
Eq.~(\ref{1D box FW and
En}) with quadratic dependence on $n$ ($E_{n} \sim n^{2}$),

\item Spectrum of the one-dimensional harmonic oscillator as in
Eq.~(\ref{Harmonic
oscillator WF and En}) with linear dependence on $n$ ($E_{n}\sim n$),

\item Intermediate spectrum that is neither of the above two types.
\end{itemize}

\begin{figure}[htbp]
\begin{center}
\leavevmode
\epsfxsize = 8.5cm 
\epsffile{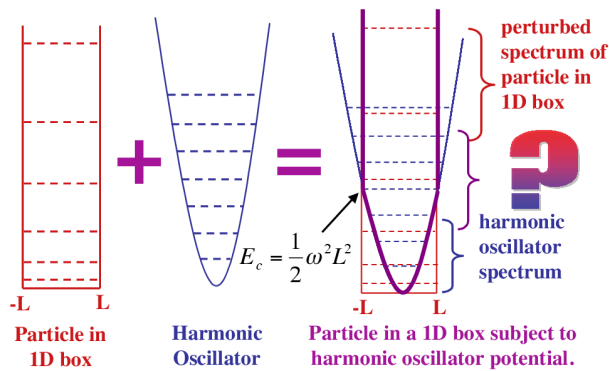}
\end{center}
\caption{Two-mode toy system consisting of a particle in a one-dimensional
box subject to a central harmonic oscillator restoring force.}
\label{1D+HO-potential}
\end{figure}

\begin{figure}[htbp]
\begin{center}
\leavevmode
\epsfxsize = 8.5cm 
\epsffile{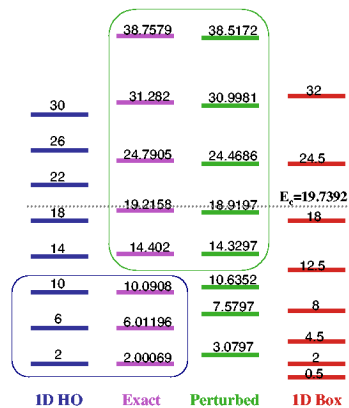}
\end{center}
\caption{Exact energies of a two--mode system with $m=\hbar =2L/\pi =1$
and $\omega =4$ compared to the spectrum of the one-dimensional harmonic oscillator (left), spectrum of the free particle in a 1D box (right), and spectrum as calculated within a first order perturbation theory of a free particle in a 1D box perturbed by a 1D HO potential. The lowest three eigenenergies of the two--mode system nearly coincide with the 1D HO eigenenergies, while higher energy states are better described as perturbations of the other limit of a free particle in a 1D box.}
\label{w4SpectralStructure}
\end{figure}

As shown in Fig.~\ref{1D+HO-potential}, one expects to see the particle in a
box spectrum at high energies. These energies correspond
to the box boundaries dominating over the harmonic oscillator
potential. In this regime, one can use standard perturbation theory to
calculate the energy for a particle in a box perturbed by a harmonic
oscillator potential. It can be shown that perturbation theory gives
better results for higher energy levels. For $n\rightarrow \infty $, the
first correction ($\delta E_{n}^{1}$) approaches a constant value:
\[
\delta E_{n}^{1}= \frac{1}{6}m\omega ^{2}L^{2}
\left( 1-\frac{6}{(1+n)^{2}\pi^{2}}\right)
\rightarrow \frac{1}{6}m\omega ^{2}L^{2}.
\]
Using $E_{n+1}^{0}-E_{n}^{0}\gg \left\langle n\left| V\right| n\right\rangle$,
one estimates that perturbation calculations are valid when
\begin{equation}
n\gg {\frac{2 m^{2}\omega ^{2}L^{4}}{3\hbar ^{2}\pi ^{2}}}.
\label{1D box spectrum begins}
\end{equation}
This analysis is confirmed by the numerical calculations shown in Fig.~\ref
{w4SpectralStructure} where the perturbed particle in a box spectrum
provides a good description for $n>6$ for the case of $m=\hbar =2L/\pi =1$ and
$\omega=4$.
Actually, the agreement extends to lower $n$ as is often the
case, perturbation theory seemingly yielding valid results well past its expected
region of validity. Note that the first order corrections are already
close to the  limiting constant value of $\frac{1}{6}m\omega^{2}L^{2}$.
On the other hand, first-order perturbation clearly fails for the ground
state. Indeed, an earlier study\cite{Armen and Rau} of the inadequacy of
fixed basis calculations showed that adequate convergence for the  ground
state when $\omega$ is large requires a large number of basis states of
the one-dimensional box. This is related to the fact that for large values
of $\omega$, or equivalently large $L$ values, a large number of 
particle-in-a-box 
wave functions ($\sin$ and $\cos$) are needed to 
obtain
the correct behavior (exponential fall off) of the low-energy harmonic
oscillator wave functions in the classically forbidden zone.

The intermediate spectrum should be observed when the harmonic oscillator
turning points coincide with the walls of the box. Therefore, the critical
energy that separates the two extreme spectral structures is given by
\begin{equation}
E_{c}=\frac{m\omega ^{2}}{2}L^{2}.
\label{Ec for 1D box and HO}
\end{equation}
Notice that the constant energy shift $m\omega ^{2}L^{2}/6$ in the energy of
the high energy levels $\delta E_{n\gg 1}^{1}$ is one-third of the critical
energy ($E_{c}/3$).

At low energies, where the classical turning points of the
oscillator lie far from the boundaries, we expect the spectrum to
coincide essentially with that of the oscillator as shown in
Fig.~\ref{w4SpectralStructure}. The number of such nearly harmonic
oscillator states is easily estimated using
\begin{equation}
E_{c}>E_{n}^{HO}\Rightarrow n_{\max }^{HO}=\frac{1}{2}\frac{m\omega L^{2}}{
\hbar }-\frac{1}{2}.
\label{HO spectrum ends}
\end{equation}
There is also a compatible number of levels,
usually larger than $n_{\max }^{HO}$, below the $E_{c}$ corresponding to a
free particle in a box,
\begin{equation}
E_{c}>E_{n}^{1D}\Rightarrow n_{\max }^{1D}=\frac{2}{\pi }\frac{m\omega L^{2}
}{\hbar }-1.
\label{1Dbox spectrum ends}
\end{equation}
However, these states are mixed by the harmonic oscillator potential.

Using the ratio of the ground state energies,
$E_{0}^{HO}/E_{0}^{1D}=4m\omega L^{2}/(\hbar \pi ^{2})$,
together with Eq.~(\ref{HO spectrum ends}) and 
Eq.~(\ref{1Dbox spectrum ends}), the following spectral situations apply:

\begin{itemize}
\item For $\left( \frac{\pi }{2}\right) ^{2}<\frac{m\omega L^{2}}{\hbar }$,
there are levels below $E_{c}$ corresponding to the harmonic oscillator and
the particle in a box such that $E_{0}^{HO}>E_{0}^{1D}.$ The 
oscillator levels
dominate the low energy spectrum.

\item For $\frac{\pi }{2}<\frac{m\omega L^{2}}{\hbar }<\left(
\frac{\pi }{2}\right) ^{2}$,
there are only the ground states $E_{0}^{1D}$ and $
E_{0}^{HO}$ below $E_{c}$ and $E_{0}^{1D}>E_{0}^{HO}$.

\item For $1<\frac{m\omega L^{2}}{\hbar }<\frac{\pi }{2}$, there is only the
ground state of the harmonic oscillator $E_{0}^{HO}$ below $E_{c}$.

\item For $0<\frac{m\omega L^{2}}{\hbar }<1$,
perturbation theory of a particle in a box should be applicable for
all levels.
\end{itemize}
The dimensionless parameter 
$\beta=\frac{m\omega L^{2}}{\hbar }$ thus plays a role similar to the
parameter $\lambda$ in Eq.~(\ref{2-mode system H}). 
In our abstract case of Eq.~(\ref{2-mode system H}), the two limits of $H$ are at $\lambda=0$ 
and $\lambda=1$ with strong mixing at $\lambda=1/2$. In the case above,
the two limits are $\beta=0$ and $\beta=\infty$ with strong mixing when  
$1<\beta<(\pi/2)^2$. In this respect, we can formally relate $\beta$ to $\lambda$ using an expression of the form $\lambda=\beta/(\beta+\beta_c)$ where $\beta_c$ is the value of
$\beta$ in the strong mixing region. For example, one may chose $\beta_c=\pi/2$.

Consider as a numerical illustration of the two-mode spectra the
case of $m=\hbar =1$, $L=\pi
/2$ and $\omega =4$ shown in Fig.~\ref{w4SpectralStructure}. With
these parameters, Eq.~(\ref{HO spectrum ends})
gives $n_{\max }^{HO}=4.53.$ Thus one should see no more than $4$
equidistant states. Indeed, in
Fig.~\ref{w4SpectralStructure}, there are four
equidistant energy levels
that correspond to a harmonic oscillator spectrum.

With respect to the critical energy $E_{c}$, there is a more explicit
classification of the spectral structure:

\begin{itemize}
\item Perturbed particle in a one-dimensional box spectrum for
energies
$E\gg E_{c}$ such that Eq.~(\ref{1D box spectrum begins}) holds,

\item One-dimensional harmonic oscillator spectrum in Eq.~(\ref{Harmonic
oscillator WF and En}) for energies $E_{c}\gg E$ such that Eq.~(\ref{HO spectrum
ends}) holds,

\item Intermediate spectrum for energies $E\approx E_{c}$.
\end{itemize}

\section{Hilbert Space of the Basis Wave Functions}
\label{Hilbert space}

\quad Before discussing the toy model using an oblique basis, it is
instructive to discuss briefly the harmonic oscillator problem in
Eq.~(\ref{Harmonic oscillator Hamiltonian}) using the wave functions for
a free particle in a one-dimensional box as in Eq.~(\ref{1D box FW
and En}); and vice versa,
solving the problem of a free particle in a one-dimensional box in
Eq.~(\ref{1D box
Hamiltonian}) using the wave functions for a particle in the harmonic
oscillator potential given in Eq.~(\ref{Harmonic oscillator WF and En}).

\begin{figure}[htbp]
\begin{center}
\leavevmode
\epsfxsize = 8.5cm 
\epsffile{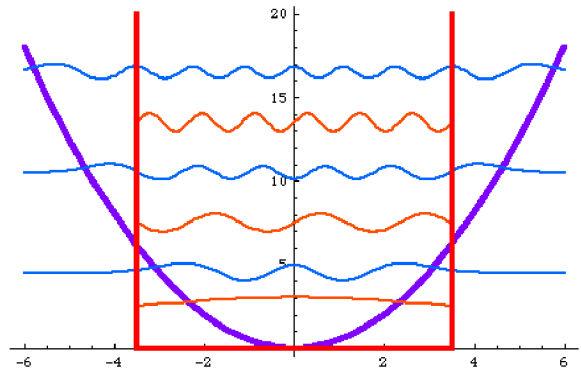}
\end{center}
\caption{Spreading of the wave functions:  
harmonic oscillator wave functions spread outside the 
harmonic oscillator potential into the classically forbidden region; 
particle in a box wave functions are zero at and outside of the box boundary. }
\label{wf-spread}
\end{figure}

Due to the different domains of the wave functions, there are some
specific problems that need to be addressed. For example, using wave
functions for a free
particle in a one-dimensional box to solve the pure harmonic oscillator problem
may not be appropriate especially for high energy states $E\gg E_{c}$.
This is because all such functions vanish outside the box (see Fig.\
\ref{wf-spread}) unlike the oscillator wave functions, especially as
the energy increases. The converse is also problematic because
oscillator functions that are non-zero outside the box will lead to
infinite energy.

The influence of the boundary conditions on the properties of a 
quantum-mechanical
system has been recognized from the dawn of 
quantum mechanics. It
is well known that some problems with seemingly separable
Hamiltonians may re-couple due to the boundary conditions.\cite{Tanner-1991}
Some recent studies on the problem of
confined one-dimensional systems using equations for relevant cut-off
functions have been discussed by Barton, Bray, and
Mackane.\cite{Barton-Bray-Mckane-1990}
Their method has been further developed in a more
general setting by Berman.\cite{Berman-1991} Other authors aim at
variational procedures using simple cut-off
functions\cite{Marin and Cruz-1991 AJP, Marin and Cruz-1991 JPB}
or derive asymptotic estimates for
multi-particle systems using the Kirkwood-Buckingham variational
method.\cite{Pupyshev and Scherbinin-1999} Somewhat different
approaches focus on
shape-invariant potentials and use supersymmetric partner potentials to
derive energy shifts and approximate wave
functions,\cite{Dutt-Mukherjee-Varshni-1995} as well as
dependence of the
ground-state energy on sample size.\cite{Monthus et al. -1996} In the 
next few paragraphs,
we discuss the structure of the relevant Hilbert spaces when confinement is
present.

\subsection{Harmonic Oscillator in the One-Dimensional Box Basis}

\quad Consider the harmonic oscillator in Eq.~(\ref{Harmonic oscillator Hamiltonian}) 
using the wave functions for a free particle in a one-dimensional box in 
Eq.~(\ref{1D box FW and En}). 
There are no practical difficulties for energies $E \ll E_{c}$ as defined in Eq.~(\ref{Ec for 1D box and HO}) as long as the turning points of the oscillator are sufficiently deep into the box (sufficiently  far from the walls of the box).
However, for energies $E\gg E_{c}$, the basis wave functions are localized only on the
interval $[-L,L]$ and thus cannot provide the necessary spread over the width of the potential (Fig. \ref{wf-spread}). This situation would be appropriate for the toy
model in Eq.~(\ref{H-ho-1Dbox}) but not for the pure harmonic oscillator problem in 
Eq.~(\ref{Harmonic oscillator Hamiltonian}).

One simple solution to the spreading problem is to continue the basis
wave functions using periodicity. This way the necessary spread of the
basis wave functions can be achieved and the new basis will stay
orthogonal but must be re-normalized. If one continues the wave functions
to infinity,  normalization will require Dirac delta functions but for
continuation on a finite interval, the functions can be normalized
to unity  as usual. However, these basis wave functions do not decay to
zero in the classically forbidden zone. This means that a significant
number of basis wave functions will be needed to account for the
appropriate behavior within the classically forbidden zone.

Another alternative is to change the support domain corresponding to
non-zero values of the function by stretching or squeezing, accomplished
through a scaling of the argument of the basis wave functions,
$x\rightarrow x\alpha _{n}/L$. This way the support becomes $[-L,L]$
$\rightarrow $ $[-\alpha _{n},\alpha _{n}]$. Here, $\alpha _{n}$ is a
scale factor for the $n$-th basis wave function in Eq.~(\ref{1D box FW
and En}), estimated either  from the width of the harmonic oscillator
potential, or determined by variational  minimization. Either way, the
new set of basis functions will be non-orthogonal. In general, there may
even be a linear dependence. However, for the basis functions discussed
here, linear dependence may not appear due to the different number of
nodes for each wave function. The number of nodes (zeros) is not changed
under the re-scaling procedure. While the potential width scaling is
simpler, its applicability is more limited than the
variationally-determined one which can be extended to more general
situations.\cite{VGG PhD}

\subsection{Particle in a Box in the Harmonic-Oscillator Basis}

\quad When the choice of the basis is not made with appropriate
care, an operator that should be Hermitian may become non-hermitian. 
In nuclear physics, although
this is unlikely for a finite shell-model calculation
using an occupation number representation,\cite{Heyde's-shell model} it is
an obstacle when one wishes to use a hard core potential and
a harmonic-oscillator basis.\cite{MoshinskyBookOnHO}

Suppose we want to solve the problem of a free particle in a one-dimensional
box $[-L,L]$ as given in Eq.~(\ref{1D box Hamiltonian}) using the
harmonic oscillator wave functions in Eq.~(\ref{Harmonic oscillator
WF and En}). The first thing to do is to change the inner product of
the wave functions:

\begin{equation}
\left(f,g\right) =\int_{-\infty
}^{\infty }f^{*}\left( x\right) g\left(
x\right) dx\rightarrow
\int_{-L}^{L}f^{*}\left( x\right) g\left( x\right) dx.
\label{inner-product-change}
\end{equation}

Then, it is immediately clear that the set of previously orthonormal
harmonic oscillator wave functions $\Psi _{n}(q)$ in
Eq.~(\ref{Harmonic oscillator WF and En}) will lose their
orthonormality and even their linear independence. The set of
functions $\Psi _{n}\left( q\right)$ with support domain restricted
to $[-L,L]$ and denoted by $\Psi _{n}(q;[-L,L])$ become linearly
dependent if $L$ is so small that there is more than one function
$\Psi _{n}(q;[-L,L])$ with the same number of nodes within
$[-L,L]$. While this linear dependence can be handled, the real
problem is the loss of hermiticity of the physically significant
operators. Neither the variational nor the potential-width wave
function scaling will help to cure the loss of hermiticity of
operators, such as those of linear momentum and energy.
This non-hermiticity is due to the behavior of the basis states at
the boundary, mainly the non-vanishing of the wave functions at $-L$
and $L$.

To understand the loss of hermiticity, we look at the off-diagonal matrix
elements of the momentum operator ($\hat{p}=-i\hbar \frac{\partial }{\partial q}$).
After some trivial manipulations, we have:
\begin{eqnarray*}
(\Psi _{m},\hat{p}\Psi _{n}) = (\hat{p}\Psi_{m},\Psi_{n}) -i\hbar \left. (\Psi
_{m}^{*}(q)\Psi _{n}(q))\right| _{-L}^{L}.
\end{eqnarray*}
It is clear from the above expression that
{\it hermiticity will be maintained only when all of the basis
functions are zero at the boundary of the interval} [-$L,L$].
Actually, the necessary condition is that the wave functions have
the same value at $\pm L$; they need to be zero only for the case of 
an infinite potential wall.

\begin{figure}[htbp]
\begin{center}
\leavevmode
\epsfxsize= 8.5cm 
\epsffile{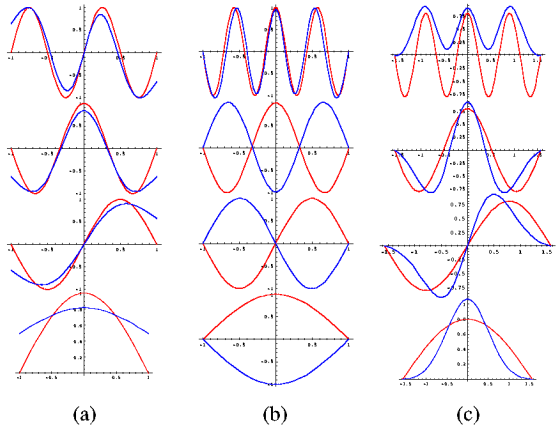}
\end{center}
\caption{Harmonic-oscillator trial wave functions (dark gray) adjusted 
with respect to the one-dimensional box problem: (a) adjusted according 
to the potential width $E_{n}^{1D}=\omega _{n}^{2}L^{2}/2\Rightarrow \omega
_{n}=\frac{ \hbar }{L^{2}}\left( 1+2n\right)$, (b) nodally adjusted
(first three are deliberately phase shifted),
(c) boundary adjusted using $\Psi (q)\rightarrow \Psi (q)-\Psi \left(
L\right) (1+q/L)/2-\Psi \left( -L\right) (1-q/L)/2$. The exact wave
functions (light gray) for a particle in a box are zero at $\pm 1$,
as clearly seen in the (a) graphs.}
\label{regularized-fw}
\end{figure}

In our simple example, all operators are built from the momentum operator $\hat p$ and position operator $\hat q$. Thus, in order to ensure hermiticity, it is sufficient to make sure that  $\hat p$ is Hermitian which requires the basis wave functions to vanish at the boundaries $-L$ and $L$. 
For this purpose, one can look at the nodes of each
basis wave function and scale it so that its outer nodes are at the
boundary points. From the nodal structure of the harmonic oscillator
wave functions, it is clear that the first two wave functions 
($\Psi_{0}$ and $\Psi _{1}$) cannot be used since they have fewer than
two nodes. Since the physical requirement that the wave functions
have to be zero at the boundary is the cornerstone in quantizing the
free particle in a one-dimensional box as in Eq.~(\ref{1D box FW and
En}), it is not surprising that the nodally adjusted harmonic
oscillator wave functions are very close to the
exact wave functions for the free particle in a one-dimensional box as shown
in Fig.~\ref{regularized-fw}.

In general, calculating the nodes of a function may become very complicated.
To avoid problems with finding the roots, one can evaluate the
wave function at the boundary points, then shift the wave function by a
constant to get zeros at the boundary, 
$\Psi (q)\rightarrow \Psi(q)-\Psi \left( L\right) $. 
This idea works well for even parity wave
functions, but has to be generalized for odd parity by adding
a linear term, 
$\Psi (q)\rightarrow \Psi (q)-(\Psi\left( L\right)
/L)q$. Thus, for a general
function, we can have: $\Psi (q)\rightarrow
\Psi (q)-(1+q/L)\Psi \left(
L\right) /2-(1-q/L)\Psi \left( -L\right)
/2$. In Fig.~\ref{regularized-fw}
we have shown some of the resulting
wave functions. Notice that this
procedure gives a new wave function
$\Psi $ that has good behavior inside
the interval $[-L,L]$ and grows
linearly with $q$ outside the interval $
[-L,L]$. This contrasts with
the behavior of the cut-off function $f(q)$ obtained by 
Barton {\it et al},\cite{Barton-Bray-Mckane-1990} where the function 
$f(q)$ has an $L/q$ singularity at the origin ($q=0$). The use of a cut-off
function to enforce boundary conditions has been studied by
Barton {\it et al}\cite{Barton-Bray-Mckane-1990} and
Berman\cite{Berman-1991} and provides an interesting integral
equation for the cut-off function. On the other hand, a simple
cut-off function supplemented by a variational method seems to
be very effective.\cite{Marin and Cruz-1991 AJP,Marin and
Cruz-1991 JPB,Pupyshev and Scherbinin-1999}

An alternative, more involved construction can be explored which relies on the Lanczos algorithm. This algorithm is an iterative process that uses the Hamiltonian of the system to generate a new basis state from the previous state at each iteration. By using the boundary matching process above, one can set up and successfully run a modification of the usual Lanczos algorithm\cite{Lanczos-1950} to
solve for the few lowest eigenvectors of the free particle in a
one-dimensional box through an arbitrary, but reasonable, choice of
the initial wave function.\cite{VGG PhD}
The major modification is to project every new function,
$\Psi_{n+1}=$ $H\Psi_{n}$,
into the appropriate Hilbert space and
subtract the components along
any previous basis vectors. Only then
should one attempt to evaluate the
matrix elements of $H$ related to
the new basis vector that is clearly
within the correct Hilbert
space. This way, one has to double the number of scalar product
operations compared to the usual Lanczos algorithm where
the matrix elements of $H$ are calculated along with the complete
re-orthogonalization of the basis vectors.

\subsection{Oblique Basis for the Two-Mode System}

\quad
In the previous two subsections we saw that a basis that is appropriate for one mode of our model system, harmonic oscillator in a box, is not appropriate for the other mode. If we desire a description in the critical mixed region characterized by  $E_c$ as shown in Fig.~\ref{1D+HO-potential}, a combination of the two basis sets seems appropriate. This is referred to as an oblique basis, stemming from a geometrical analogy. A two-dimensional space is usually described in terms of the Cartesian $(x,y)$ axes or, indeed, in terms of any other orthogonal pair obtained by rotating those axes. While such orthogonal choices are more convenient and familiar, any two axes, as long as they are not linearly dependent, also suffice to describe the space. Such a choice constitutes an ``oblique" pair. Similarly, when we mix both harmonic oscillator and particle in a box states, we have an oblique basis. 

In using such an oblique basis, there are two main problems to be addressed in order to have a proper Hilbert space of our quantum mechanical system. 
First, we have to make sure that any set of states that are derived from harmonic oscillator functions satisfies the boundary conditions of the problem. For the particle in a box states the boundary conditions are satisfied by construction. We discussed already a few possible ways to construct states with the correct boundary conditions. An interesting  additional method, suggested by one of our referees, would be to use Hermite functions with non-integer index. For such functions with index between $n$ and $n+1$, the position of the outer node is correspondingly between the outer nodes of the $n$-th and $(n+1)$-th Hermite functions. Second, after the chosen set of functions has been modified to satisfy the boundary conditions, the orthonormality of these functions would most likely be destroyed. Even if the two basis sets are orthonormal by themselves, they would not be orthonormal as a whole and may even be linearly dependent. While this might seem a serious technical problem, it has a well known solution through re-orthonormalization of the basis or by proceeding with a generalized eigenvalue problem.\cite{VGG 24MgObliqueCalculations}  

Our oblique basis consists of modified harmonic oscillator (MHO) basis  states that satisfy the boundary conditions along with basis states of a free particle in a box (BOX). To obtain the wave functions and eigenenergies we solve the generalized eigenvalue problem within this oblique basis.  Schematically, these basis vectors and their overlap matrix can be represented in the following way:
\begin{eqnarray}
{\rm basis ~vectors} &:& {\cal E}=\left(
\begin{array}{c}
e_{\alpha}:{\rm box - ~basis} \nonumber\\
e_{i} :{\rm mho - ~basis}
\end{array}
\right), \label{Basis vectors} \nonumber\\
{\rm overlap ~matrix} &:& \Theta =\left(
\begin{array}{lr}
{\bf 1} & \Omega \\
\Omega^{+} & {\mu}
\end{array}
\right)
\begin{array}{c}
 ~\Omega_{\alpha i} = e_{\alpha }\cdot e_{i},\\
 ~\mu_{i j} = e_{i }\cdot e_{j},
\end{array}
\label{Overlap matrix} \nonumber\\
{\rm hamiltonian} &:& H=
\left(
\begin{array}{ll}
H_{\alpha \beta } & H_{\alpha j} \\
H_{i\beta } & H_{ij}
\end{array}
\right)=\nonumber\\
&~& \quad= \left(
\begin{array}{ll}
H_{box \times box} & H_{box \times mho} \\
H_{mho \times box} & H_{mho \times mho}
\end{array}
\right),\nonumber
\label{hamiltonian Matrix}
\end{eqnarray}
where $\alpha=1, ...,$ dim(box -- basis) and 
$i =1, ...,$ dim(mho -- basis).

In these notation, the eigenvalue problem: 
$$H v=E v$$ 
with
$$v = v^\alpha e_\alpha+v^i e_i $$
takes the form:
\begin{eqnarray}
\left(\begin{array}{ll}
H_{\alpha \beta } & H_{\alpha j} \\
H_{i\beta } & H_{ij}
\end{array}\right)
\left(\begin{array}{c}v^{\beta} \\ v^{j}\end{array}\right) &=& E
\left(\begin{array}{lr} {\bf 1} & \Omega \\ \Omega^{+} & {\mu} \end{array}\right)
\left(\begin{array}{c}v^{\beta} \\ v^{j}\end{array}\right).\nonumber
\end{eqnarray}
which is a generalized eigenvalue problem.

\section{Discussion of the Toy Model Calculations and Results}
\label{Calculations and Results}

\quad Despite the simplicity of the toy model in
Eq.~(\ref{H-ho-1Dbox}), the
harmonic oscillator in a box exhibits
some of the essential characteristics
of a more complex system. Some
of our interest lies in problems associated
with the use of
fixed-basis calculations. In particular, one such problem is the slow
convergence of the calculations.\cite{Armen and Rau} If one can
implement an exact arithmetic, one may not worry too much about the slow
convergence when enough time, storage, and other computer resources
are provided. However, numerical calculations are plagued with
numerical errors so that
a calculation that converges slowly may be
compromised by accumulated
numerical error. Of course, having the
correct space of functions is of the
utmost importance in any calculation.

Having considered the main problems one may face in
studying the simple toy
model in Eq.~(\ref{H-ho-1Dbox}),
we continue
our discussion with the spectrum for the case of $m=\hbar
=2L/\pi =1$
and $\omega =4$.
As one can see in Fig.~\ref{w4SpectralStructure},
the first three energy
levels are equidistant and almost
coincide with the harmonic oscillator
levels as expected from
Eq.~(\ref{HO spectrum ends}). For these
states, the wave functions
are also essentially the harmonic oscillator wave functions. The
intermediate spectrum is almost missing. Above $E_{c}$, the spectrum
is that of a free particle in a one-dimensional box perturbed by the
harmonic oscillator potential. We find that an oblique-basis calculation
reproduces the lowest eight energy levels using 14 basis functions,
seven nodally adjusted harmonic oscillator states and
seven states of a free particle in a box. In contrast, a
fixed-basis calculation, using only the wave functions of a free
particle in a one-dimensional box, requires 18 basis states.

Due to the simplicity of the toy model, it does not appear as if the
oblique-basis calculation has a big numerical advantage over
calculations using the fixed basis of the box wave functions. There
are two main reasons for this: (1) there is a critical energy
$E_{c}$ that separates the two modes, (2) the spectrum above 
$E_{c}$ 
has a nice regular structure.

The regular structure
above $E_{c}$ results in a very
favorable situation for
the usual fixed-basis calculations since the
dimension of the space
needed to obtain the $n$-th eigenvalue grows
as $n+\alpha$. The
parameter $\alpha$ is relatively small and does
not change much in a
particular region of interest. For example, the
$\omega =16$
calculations need only $\alpha =15$ extra basis vectors when
calculating any of the eigenvectors up to the hundredth. The
relatively constant value of $\alpha $ can be understood by
considering the harmonic oscillator potential as an interaction that
creates excitations out of the $n$-th
unperturbed box state.
Therefore, $\alpha $ is the number of box
states with energies in the
interval $E_{n}^{0}$ and $E_{n}^{0}+\omega
^{2}/2\left\langle \Phi
_{n}\right| x^{2}\left| \Phi _{n}\right\rangle $
where $E_{n}^{0}$ is
the $n$-th unperturbed box state energy. There is a
rapid de-coupling
of the higher energy states from any finite excitation
process that
starts out of the $n$-th state. This de-coupling is due to the
increasing energy spacing of the box spectrum which results in a
limited number of states mixed by the harmonic
oscillator
potential. Using the upper limit $E_{c}/3$ on $\delta
E_{n}^{1}$, one can
estimate $\alpha \approx
\frac{1}{\sqrt{3}}n_{\max }^{1D}$.

The sharp separation of the two
modes also allows for a safe use of the harmonic oscillator states
without any rescaling. This is especially true when $\omega$ is very
large since then the low energy states are naturally localized within
the box. Therefore, instead of diagonalizing the Hamiltonian in a box
basis, one can just use the harmonic oscillator wave
functions.

\begin{figure}[htbp]
\begin{center}
\leavevmode
\epsfxsize= 8.5cm 
\epsffile{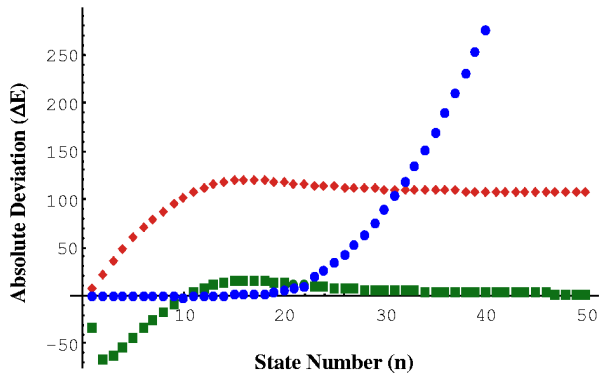}
\end{center}
\caption{Absolute
deviations of variously calculated energies from the exact energy
eigenvalues for $\omega=16$, $L=\pi /2$, $\hbar =m=1$ as a function
of $n$. Circles represent deviation of the exact energy eigenvalue
from the corresponding harmonic oscillator eigenvalue ($\Delta
E=E^{exact}_n-E^{HO}_n$), the diamonds are the corresponding
deviation from the energy spectrum of a particle in a box ($\Delta
E=E^{exact}_n-E^{1D}_n$), and the squares are the first-order
perturbation theory
estimates.}
\label{DeltaEn}
\end{figure}

\begin{figure}[htbp]
\begin{center}
\leavevmode
\epsfxsize = 8.5cm 
\epsffile{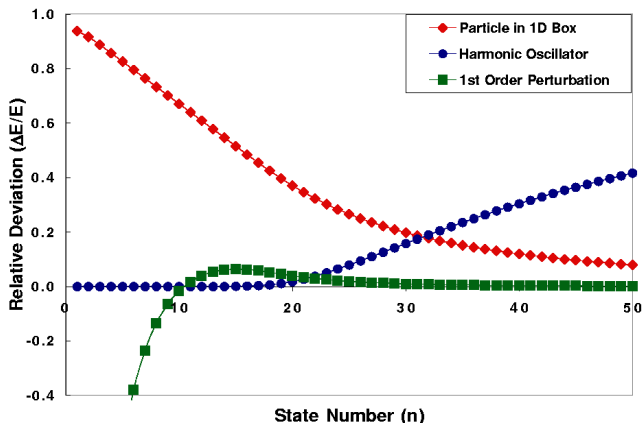}
\end{center}
\caption{As in
Fig.~\ref{DeltaEn} but for relative deviations from
the exact energy
eigenvalues of the three
calculations.}
\label{DeltaEoverE}
\end{figure}

Fig.~\ref{DeltaEn} shows the absolute deviation
($\Delta E=E_{n}^{exact}-E_{n}^{estimate}$)
of the calculated energy spectrum for the case
of $\omega =16$, $L=\pi /2$, $\hbar =m=1$. Here, $E_{n}^{estimate}$
stands for one of the three energy estimates one can make: the
harmonic oscillator $E^{HO}_n$, particle in a one-dimensional box
$E^{1D}_n$, and the
first order perturbation theory estimate considering the harmonic
oscillator potential as a perturbation,
($E_{n}^{1D}+\omega^{2}/2\left\langle \Phi _{n}\right|
x^{2}\left| \Phi_{n}\right\rangle$).
There are about 19 states that match the harmonic oscillator
spectrum which is consistent with the expected value from
Eq.~(\ref{HO spectrum ends}). After the $n=20$ level, perturbation
theory gives increasingly better results for the energy eigenvalues.
Fig.~\ref{DeltaEoverE} shows the relative deviation
($1-E_{n}^{estimate}/E_{n}^{exact}$).

\begin{figure}[htbp]
\begin{center}
\leavevmode
\epsfxsize = 8.5cm 
\epsffile{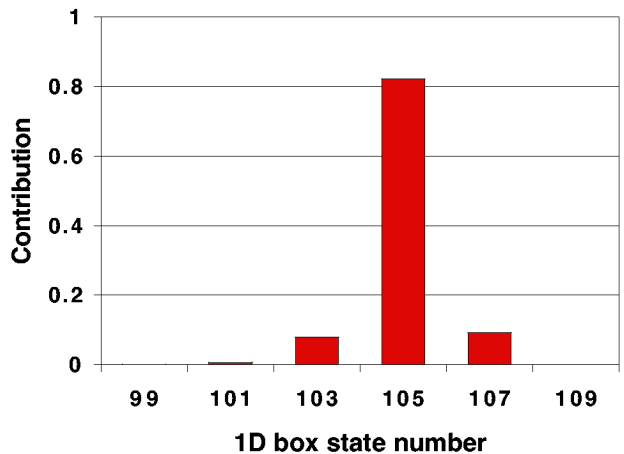}
\end{center}
\caption{Non-zero
components of the 105th exact eigenvector in the basis of
a free particle in a one-dimensional box. Parameters of the
Hamiltonian are
$\omega =16$, $L=\pi /2$, $\hbar
=m=1$.}
\label{State105}
\end{figure}

Note that
perturbation theory is valid, as expected, for high energy states
determined by Eq.~(\ref{1D box spectrum begins}). For the
high energy
spectrum, the harmonic oscillator potential acts as a small
perturbation. Thus the first-order corrections in the energy
and the wave
function are small. Fig.~\ref{State105} shows that the
main component of the
$105^{th}$ exact wave function comes from the
$105^{th}$ box wave function, as it
should in a region of small
perturbations.

\begin{figure}[htbp]
\begin{center}
\leavevmode
\epsfxsize= 8.5cm 
\epsffile{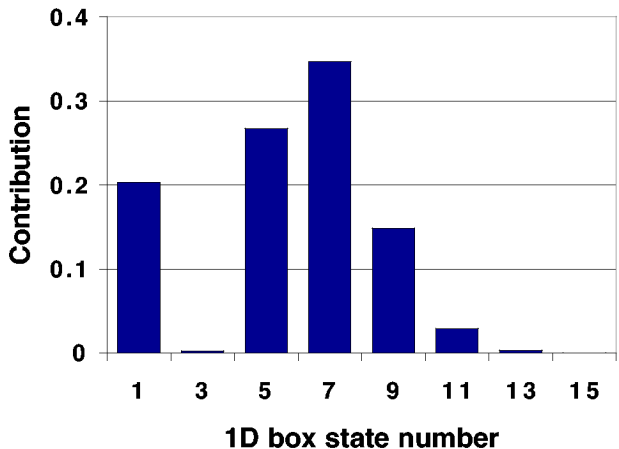}
\end{center}
\caption{Non-zero
components of the third harmonic oscillator (even parity) eigenvector as
expanded in the basis of a free particle in a one-dimensional box.
Parameters of the Hamiltonian are $\omega =16$, $L=\pi /2$, $\hbar
=m=1$.}
\label{State3}
\end{figure}

For low energy states,
perturbation theory around the box states is not
appropriate, the
harmonic oscillator states being closer to the true states in this
region. Specifically, for $m=\hbar =2L/\pi =1$ and $\omega =16,$
the first ten states are essentially the harmonic oscillator states
to a very high accuracy. The next ten states have still high overlaps
with the corresponding
harmonic oscillator wave functions. For
example, starting from 0.999999 at
the tenth state, the overlaps go
down to 0.880755 at the twentieth state.
After that the overlaps get
small very quickly. Fig.~\ref{State3} shows the
structure of the
third exact eigenvector when expanded in the box basis.
Notice that
the third box wave function is almost missing. An
explanation lies in the
structure of the harmonic oscillator
functions, which are
essentially exact in this region. Upon
projecting these functions in
Eq.~(\ref{Harmonic
oscillator WF and En}) onto the box functions in
Eq.~(\ref{1D box FW and
En}), the results can be obtained in closed
analytical form. The
integrand consists of three factors, an even
power of $q$, a Gaussian
and a cosine. While the oscillations of the
Hermite polynomial are of
varying amplitude, the cosine has evenly
spaced nodes and antinodes
of equal amplitude. As a result,
cancellations can take place between
successive terms in the
integrand. For the third oscillator function
(second of even
parity), its one pair of nodes is reflected in the
dip seen at $n=3$
in Fig.~\ref{State3}. Higher oscillator functions
with more pairs of
nodes can display correspondingly more such dips
in the projected squared amplitudes.

\begin{figure}[htbp]
\begin{center}
\leavevmode
\epsfxsize = 8.5cm 
\epsffile{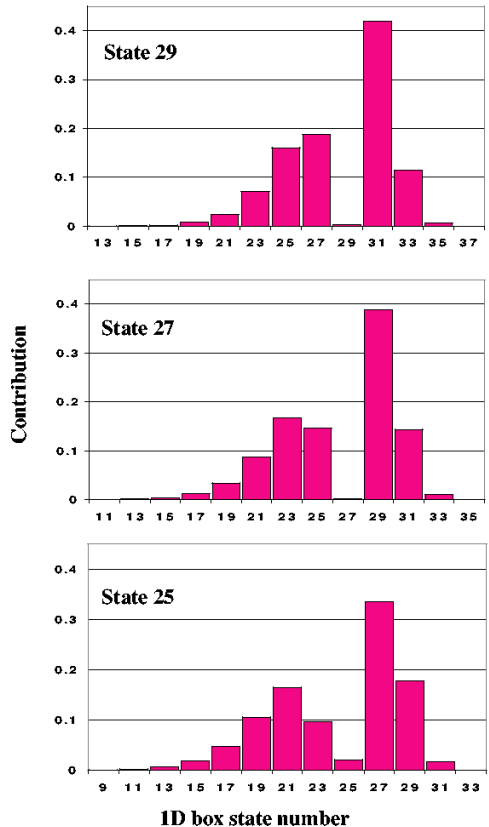}
\end{center}
\caption{Coherent structure with respect to the non-zero components of the
25th, 27th and 29th exact eigenvector in the basis of a free particle in a
one-dimensional box. Parameters of the Hamiltonian are $\omega =16$, $L=\pi/2
$, $\hbar =m=1$.}
\label{States25to29}
\end{figure}

This pattern of having a small
projection of the exact wave function along
the corresponding box
wave function continues to persist into the
transition region. This
may seem unexpected considering the fact that
the first order
estimates of the energy levels are relatively good.
Notice that in our two examples of $\omega=4$
(Fig.~\ref{w4SpectralStructure}) and
$\omega=16$ (Fig.~\ref{DeltaEn}), the first order corrections in the
transition region are already close to the limiting constant value of
$\frac{1}{6}m\omega^{2}L^{2}$, even though the corresponding
box wave functions are not present at all in the exact wave function
as shown in Fig.~\ref{States25to29}.

From the results in these graphs, it seems that the transition
region is absent
since first-order perturbation theory becomes valid immediately after the
breakdown of the harmonic oscillator spectrum. That first-order
perturbation theory gives good estimates for the energy levels in this
transition region is a manifestation of coherent behavior. What actually
happens in this region is a coherent mixing of box states by the harmonic
oscillator potential in the sense of a quasi-symmetry.\cite{VGG PhD, VGG
SU(3)andLSinPF-ShellNuclei, Adiabatic mixing} This coherent mixing
is illustrated in Fig.~\ref{States25to29}, where one can see that the
histograms for a few consecutive states are very similar. In this
sense we say that the corresponding particle in a box states are
coherently mixed. A more precise and detailed discussion of coherent structure, 
behavior, and quasi-structures can be found elsewhere.\cite{VGG PhD}

\section{Conclusions}
\label{Conclusions}

In summary, we have studied the simplest two-mode system consisting of a one
dimensional harmonic oscillator in a one-dimensional box. Depending on the
parameters of the two exactly solvable limits, one observes various spectral
structures. There is a clear coherent mixing in the transition region. It is
remarkable that such a simple system can exhibit coherent behavior similar
to the one observed in nuclei. There is clear advantage to using
an oblique-basis set which includes both oscillator and particle in a box states.
This allows one to use the correct wave functions in
the relevant low and high energy regimes relative to $E_c$. Taking into
account the importance of the relevant energy scale of a problem and the
wave function localization with respect to the range of the potential, the
oblique-basis method can be extended beyond the idea of using two or more
orthonormal basis sets. Specifically, one can consider a
variationally-improved basis set by starting with some initially guessed
basis states. In the occupation number representation (Fock space), which
is often used in the nuclear shell model,\cite{Heyde's-shell model} this
variationally-improved basis method seems inapplicable. But
the method is of general interest because of its possible relevance
to multi-shell {\it ab-initio} nuclear physics, atomic physics, and general
quantum mechanical calculations. The method may also be related to
renormalization-type techniques.\cite{VGG PhD}

\section{Student exercises}
\label{Student exercises}

\begin{enumerate}

\item Section II views the two Hamiltonians in Eq.~(\ref{1D box Hamiltonian}) and Eq.~(\ref{Harmonic oscillator Hamiltonian}) as limits of Eq.~(\ref{H-ho-1Dbox}) for suitable choices of parameters. Express this feature instead in the form of Eq.~(\ref{2-mode system H}) with suitable definition and choice of $\lambda$ and  appropriate modification of $\omega$ in Eq.~(\ref{Harmonic oscillator Hamiltonian}).

\item Treating the harmonic oscillator potential as a perturbation, work out the first order correction to the energies of particle in a box states. Hence, verify Eq.~(\ref{1D box spectrum begins}).

\item Project the third harmonic oscillator wave function (second even parity state with two nodes) onto the wave functions of a particle in a box to verify the structure shown in Fig.~\ref{State3}. Integrals involved have products of powers of $q$, a Gaussian and a cosine in $q$, and you may take integration limits to $\pm \infty$.

\item As a variational counterpart of the oblique basis concept in this paper, the product of the ground state functions of a particle in a box and an oscillator may be chosen as a trial wave function. With such a form, calculate through the Rayleigh-Ritz variational principle the ground state energy of the system.      

\end{enumerate}

\section*{ACKNOWLEDGMENT}
This work was supported by the U.S. National Science Foundation under
grants No. PHY 0140300 and PHY 0243473, and the Southeastern Universities
Research Association. One of us (A.R.P.R) thanks the Research School of
Physical Sciences and Engineering at the Australian National University for
its hospitality during the writing of this paper. This work is partially
performed under the auspices of the U. S. Department of Energy by the
University of California, Lawrence Livermore National Laboratory under
contract No. W-7405-Eng-48.

\end{document}